# Moderate Point: Balanced Entropy and Enthalpy Contributions in Soft Matter

Baoji He (贺宝记),[1,2] Yanting Wang (王延颋),[1,2,a]

[1]*CAS Key Laboratory of Theoretical Physics, Institute of Theoretical Physics, Chinese Academy of Sciences, 55 East Zhongguancun Road, P. O. Box 2735, Beijing, 100190 China*

[2]*School of Physical Sciences, University of Chinese Academy of Sciences, 19A Yuquan Road, Beijing, 100049 China*

Various soft materials share some common features, such as significant entropic effect, large fluctuations, sensitivity to thermodynamic conditions, and mesoscopic characteristic spatial and temporal scales. However, no quantitative definitions have yet been provided for soft matter, and the intrinsic mechanisms leading to their common features are unclear. In this work, from the viewpoint of statistical mechanics, we show that soft matter works in the vicinity of a specific thermodynamic state named moderate point, at which entropy and enthalpy contributions among substates along a certain order parameter are well balanced or have a minimal difference. Around the moderate point, the order parameter fluctuation, the associated response function, and the spatial correlation length maximize, which explains the large fluctuation, the sensitivity to thermodynamic conditions, and mesoscopic spatial and temporal scales of soft matter, respectively. Possible applications to switching chemical bonds or allosteric biomachines determining their best working temperatures are also discussed.

## I. INTRODUCTION

Named by Pierre-Gilles de Gennes in his Nobel Prize lecture,[1, 2] soft matter vaguely refers to a wide spectrum of condensed materials, ranging from polymers, complex liquids, liquid crystals to foams, colloids, gels, granular materials, and biological materials. Various soft materials share some common features, such as significant entropic effect, large fluctuations, sensitivity to thermodynamic conditions, mesoscopic characteristic spatial and temporal scales, and the ability of self-assembly.[3] Despite successes in some individual fields, such as polymers[4] and liquid crystals,[5] no general quantitative definitions have yet been provided for soft matter, and the intrinsic mechanisms leading to their common features are unclear.

Regarding the subtle balance between entropy and enthalpy contributions in soft matter, the term *entropy-enthalpy compensation* has been frequently carried out since 1950s, aiming to quantitatively describe the unique characteristics of soft matter (see, e.g., Refs. 6-12). Initially, entropy-enthalpy compensation was referred to the phenomenon that a change in thermodynamic condition leads to almost identical changes in entropy and enthalpy, for the Helmholtz free energy

---

$\Delta F = \Delta U - T\Delta S \approx 0$ or for the Gibbs free energy $\Delta G = \Delta H - T\Delta S \approx 0$, where $T$ is the system temperature, $U$ is the system total energy, $H$ is the system enthalpy, and $S$ is the system entropy. In the term "entropy-enthalpy compensation", "entropy" refers to $T\Delta S$, the entropic effect caused by a finite temperature and/or finite spatial degrees of freedom, and "enthalpy" refers (not rigorously) to interactions among particles, either $U$ or $H$, for the sake of simplicity. However, this interpretation was doubted very likely an artifact caused by inappropriate data analysis methods.[13, 14] Recently, the entropy-enthalpy compensation has taken another form of

$$\Delta\Delta F = \Delta\Delta U - T\Delta\Delta S \approx 0 \qquad (1)$$

or

$$\Delta\Delta G = \Delta\Delta H - T\Delta\Delta S \approx 0. \qquad (2)$$

The physical interpretation of this form is that the free energy cost $\Delta F$ or $\Delta G$ for a certain statistical process remains roughly unchanged when the thermodynamic condition changes, attributed to the compensated entropy change with respect to the enthalpy change. Another term *entropy-enthalpy balance* vaguely refers to the phenomenon that entropy and enthalpy have comparable contributions under a certain thermodynamic condition (see, e.g., refs. 15-20), yet frequently it takes the same expression as Eq. (1) or (2).

The current forms of the above two terms, however, neither reveal the statistical physics essence of soft matter nor provide a systematic explanation of the common features soft materials exhibit. The problem may reside in the perspective of thermodynamic states for looking into the subtle balance between thermodynamic entropy and enthalpy. Instead, if we look at the substates of a single thermodynamic state, they can generally be regarded as having various ratios of entropy and enthalpy contributions, which are different from thermodynamic variables of entropy or enthalpy. In this paper, by looking into the substates of the free energy landscape along a predefined order parameter, we will define soft matter as the materials working in the vicinity of a *moderate point*, at which entropy and enthalpy contributions among substates are well balanced (or have a minimal difference when the exact balanced condition cannot be satisfied). This (quasi-)balanced condition leads to the maximization of the order parameter fluctuation around the moderate point, which explains the intrinsic large fluctuation of soft matter. The associated response function also maximizes around the moderate point, which explains the sensitivity of soft materials to thermodynamic conditions. Their mesoscopic temporal and spatial scales may be explained by the maximization of the finite spatial correlation length around the moderate point. For convenience, we will call our theory *the moderation theory*.

## II. THEORY

From now on, we assume that by default the system is in the canonical ensemble, and "enthalpy" actually refers to the system total energy. The essential idea of the moderation theory is that the subtle balance between the entropy and enthalpy contributions of soft matter should be reflected by the probability distribution (or equivalently the free energy landscape) of all *substates* of a thermodynamic state (a state point in the phase space). With an appropriately defined order parameter, a thermodynamic system with very high dimensions (e.g., a system with $N$ particles in the 3-D space has $6N$ degrees of freedom) is projected into a reduced phase space spanned by the order parameter. Note that here "substates" are microstates in the reduced phase space, not in the original full-dimensional space. For simplicity, in this paper, only the case with one order parameter $h$ is considered, along with a generalized field $B$ conjugated with $h$. Nevertheless, it should be straightforward to generalize the theory to the cases with multiple order parameters.

### A. Free energy landscape

The Helmholtz free energy of a thermodynamic state point is

$$F(\beta, B) = -\frac{1}{\beta} \ln Z(\beta, B), \tag{3}$$

where $\beta \equiv \frac{1}{k_B T}$ with $k_B$ the Boltzmann constant, and $Z$ is the total partition function. In the reduced phase space spanned by an order parameter $h$, a substate has the probability density of

$$f(h; \beta, B) = Z(h; \beta, B) / Z(\beta, B), \tag{4}$$

where $Z(h; \beta, B) = g(h; B) \exp(-\beta U(h; \beta, B))$ is the partial partition function with $g(h; B)$ the density of states and $U(h; \beta, B)$ the energy. The free energy of a substate can be defined as

$$F(h; \beta, B) = -\frac{1}{\beta} \ln Z(h; \beta, B). \tag{5}$$

Note that $Z(\beta, B) = \int_{-\infty}^{+\infty} Z(h; \beta, B) dh$ and $\int_{-\infty}^{+\infty} f(h; \beta, B) dh = 1$ but $F(\beta, B) \neq \int_{-\infty}^{+\infty} F(h; \beta, B) dh$. The *free energy landscape* of a state point along this order parameter is a collection of the free energies of all its substates, which is equivalent to the *potential of mean force* defined in chemistry. Note that $B$ is conjugate to the order parameter $h$, and does not necessarily have to be a real external field.

## B. Two-substate model

Without considering the generalized field $B$, for the simplest case of a classical statistical system with only two substates: entropy-dominated state 1 and enthalpy-dominated state 2 after projecting into the phase space spanned by an order parameter $h$, if the system energies of the two substates are $U_1$ and $U_2$ with $U_1 > U_2$, the densities of states are $g_1$ and $g_2$ with $g_1 > g_2$, and the associated order parameters are $h_1$ and $h_2$, then the system partition function is given by $Z(\beta) = Z_1(\beta) + Z_2(\beta)$, where $Z_i(\beta) = g_i \exp(-\beta U_i)$, $i = 1$ or 2, and the appearance probability $P_i(\beta) = Z_i(\beta)/Z$. The free energies associated with the two substates are $F_i = -\frac{1}{\beta}\ln Z_i = U_i - \frac{1}{\beta}\ln g_i$, $i = 1$ or 2. At low temperatures, $F_1 > F_2$, the system is enthalpy dominated; at high temperatures, $F_1 < F_2$, the system is entropy dominated. Therefore, it is natural to require

$$F_1 = F_2 \tag{6}$$

at the *moderate point* when the system has the entropy and enthalpy contributions well balanced, which determines the temperature at the moderate point

$$T_m = \frac{U_2 - U_1}{k_B (\ln g_2 - \ln g_1)}. \tag{7}$$

If the generalized field $B$ is considered with $U_1(B) > U_2(B)$ and $g_1(B) > g_2(B)$, in some cases the two substates cannot have exactly balanced entropy and enthalpy contributions when varying $B$ at a fixed temperature. Therefore, instead of requiring $F_1 = F_2$, we now define the moderate point as the one satisfying a looser condition:

$$|F_1(B_m) - F_2(B_m)| = F_d, \tag{8}$$

where $F_d \geq 0$ is the smallest free-energy difference taken at the moderate point $B = B_m$. To differentiate, the point satisfying $F_d = 0$ (i.e., $F_1 = F_2$) will be called the *exact moderate point*, otherwise it will be called the *general moderate point*, and the corresponding conditions are the *exact moderate condition* and the *general moderate condition*, respectively.

In Appendix A, we prove that the order parameter fluctuation $D_h \equiv \langle h^2 \rangle - \langle h \rangle^2$ maximizes at either the exact moderate point or the general moderate point for the two-substate model, where $\langle h \rangle = P_1 h_1 + P_2 h_2$ and $\langle h^2 \rangle = P_1 h_1^2 + P_2 h_2^2$.

## C. Continuous case

When the substates are continuous, it is unobvious what kind of probability distributions have balanced entropy and enthalpy contributions. Our strategy is dividing the whole probability distribution into two parts by the ensemble average (expected value) of the order parameter $\langle h \rangle$: one side is entropy-dominated and the other side is enthalpy-dominated. As derived in Appendix B, based on the exact moderate condition for the two-substate model given by Eq. (6), we obtain the exact moderation condition for the continuous case

$$\langle h \rangle = h_{\mathrm{m}}, \tag{9}$$

where $\langle h \rangle = \int_{-\infty}^{+\infty} h f(h;\beta,B) \mathrm{d}h$ and $h_{\mathrm{m}}$ is the median satisfying $\int_{-\infty}^{h_{\mathrm{m}}} f(h;\beta,B) = 0.5$. The general moderate condition is

$$\left| \int_{h_{\mathrm{m}}}^{\langle h \rangle} f(h) \mathrm{d}h \right| = P_{\mathrm{d}}, \tag{10}$$

where $P_{\mathrm{d}} \geq 0$ is the smallest possible deviation of the integral from 0. Note that when the difference between the entropy and enthalpy contributions is too large, it is unreasonable to still treat the system as a soft material. Nevertheless, usually the difference varies continuously with thermodynamic variables (e.g., temperature), and there are no natural boundaries distinguishing the "soft" region from other cases, a problem similar to the definition of the exact glass transition point.

As shown in Appendix C, the maximization of the order parameter fluctuation $D_h$ with respect to the generalized field $B$ has to satisfy

$$\left\langle \left( h - \langle h \rangle \right)^3 \right\rangle = 0. \tag{11}$$

Under the assumption that the system energy $U$ linearly depends on $h$, the condition for the maximization of $D_h$ with respect to the temperature $T$ (or equivalently $\beta$) is also Eq. (11). In Appendix D, we prove that the moderate condition Eq. (10) naturally leads to Eq. (11). That is, the order parameter fluctuation $D_h$ maximizes at either the exact moderate point or the general moderate point for the continuous case.

## D. Response function and spatial correlation length

In the linear response regime, the system Hamiltonian under a generalized field $B$ can be written as

$$U(h;\beta,B) = U_0(h;\beta) - Bh, \tag{12}$$

where $U_0$ is independent of $B$, and the probability density function should have the form

$$f(h;\beta,B) = g(h)\exp(\beta Bh - \beta U_0(h;\beta))/Z(\beta,B). \tag{13}$$

Since the total free energy is $F = -\frac{1}{\beta}\ln Z$, the ensemble average of the order parameter

$$\langle h \rangle = -\left(\frac{\partial F}{\partial B}\right)_\beta. \tag{14}$$

The response function corresponding to $B$ is

$$\chi \equiv \left(\frac{\partial \langle h \rangle}{\partial B}\right)_\beta = \beta D_h, \tag{15}$$

where the order parameter fluctuation $D_h \equiv \langle h^2 \rangle - \langle h \rangle^2$ usually depends on both $\beta$ and $B$.

According to Eq. (15), when varying the generalized field $B$ at a fixed $\beta$, the response function maximizes exactly at the point $D_h$ maximizes. When varying the temperature $T$ (equivalently $\beta$), this relation does not hold, but numerically the maximization of the response function appears at a point very close to the maximization of $D_h$ if $D_h$ varies slowly around its maximal point, as demonstrated in Appendix E.

The characteristic spatial scale can be quantified by the spatial correlation length. In the linear response regime, the spatial correlation length can be related to the response function as

$$\frac{1}{\beta}\chi(r) \sim \langle \delta h(0) \delta h(r) \rangle, \tag{16}$$

where $r$ is the spatial interval. In Appendix F, we show that the spatial correlation length maximizes when $D_h$ maximizes. On the other hand, no quantitative connections can be established between $D_h$ and the time correlation length, but as a rule of thumb, in view of the vast scale, for material systems, a larger spatial scale roughly corresponds to a longer temporal scale. This might explain why soft materials generally have mesoscopic characteristic spatial and temporal scales, since microscopic scales correspond to short correlation lengths and macroscopic scales correspond to divergent correlation lengths, while soft materials work under a thermodynamic condition in between.

## III. EXAMPLE of POLYGLUTAMINE AGGREGATION

The validity of the moderation theory was examined with a simplified polypeptide aggregation model. Previously, we simulated by coarse-grained molecular dynamics the aggregation behaviour of polyglutamine molecules in aqueous solution.[21, 22] We found that, in equilibrium, the instantaneous configurations of polyglutamine molecules fluctuate from distributing almost uniformly to aggregating very tightly, and the degree of aggregation decreases monotonically with temperature but first increases and then decreases with concentration. Based on the simulation data, we then projected this high-dimensional system into a one-dimensional phase space spanned by the so-called heterogeneity order parameter (HOP)[23] characterizing the degree of aggregation, defined as

$$h = \frac{1}{N} \sum_{i=1}^{N} \sum_{j=1}^{N} \exp(-r_{ij}^2 / \sigma^2), \quad (17)$$

where $N$ is the number of polyglutamine molecules, $r_{ij}$ is the distance between molecule $i$ and molecule $j$ corrected with the periodic boundary condition, and $\sigma = L/N^{1/3}$ with $L$ the side length of the cubic simulation box. A larger value of HOP represents a higher degree of aggregation. Both the density of states $g$ and the potential energy $U_p$ are represented by a function of $h$ and $L$ (directly corresponding to concentration). For molecular systems with the Newtonian dynamics, the partition function is $Z = Z_k Z_p$, where $Z_k$ is the partial partition function for the momentum space and

$$Z_p(\beta, L) = \int g(h; L) \exp(-\beta U_p(h; L)) dh. \quad (18)$$

Since the momentum space is independent of the configurational space, $Z_k$ is always cancelled out and only $Z_p$ should be considered during normalization.

For the system with 27 32-residue polyglutamines, by fitting the simulation data, we determined the density of states

$$g(h; L) = g_0 h^3 (h-14)^6 \exp\left(-\left(a_1 + a_2/L^4\right)h\right) \quad (19)$$

and the potential energy

$$U_p(h; L) = -\frac{a_3}{L} h + U_0, \quad (20)$$

where $h \in [0,14]$, $a_1 = 5.8$, $a_2 = 4.14 \times 10^4$ nm$^4$, $a_3 = 2.660$ nm·eV, and $g_0$ and $U_0$ are undetermined constants. The probability for a certain $h$ to appear is therefore

$$P(h;\beta,L) = g(h;L)\exp(-\beta U_p(h;L))/Z_p(\beta,L). \tag{21}$$

The generalized field associated with $h$ is

$$B = \frac{a_3}{L} - \frac{a_1}{\beta} - \frac{a_2}{\beta L^4}, \tag{22}$$

so that the configurational partition function can be written as

$$Z_p = \int_0^{14} g_0 h^3 (h-14)^6 \exp(\beta B h - \beta U_0) dh. \tag{23}$$

In this model, the fluctuation of the order parameter is a function of both temperature and concentration $D_h(\beta,L)$. In Figure 1a, the red line depicts the thermodynamic states at which $D_h(\beta,L)$ takes the largest possible value of 4.1869. It is interesting to see that, below the critical temperature $T_c = 316.4$ K, each $T_m$ has two corresponding $L$ values, resembling the liquid-vapour phase transition of a finite-size system. For temperatures above $T_c$, $D_h$ cannot be as large as 4.1869, and the thermodynamic states with the largest $D_h$ are drawn with a green line. In the same plot, the thermodynamic states determined by Eq. (10) are drawn with black lines, which almost perfectly overlap with the red and green lines, indicating that the order parameter fluctuations maximize at the moderate points. As shown in Figure 1b and 1c, below $T_c$, the entropy and enthalpy contributions are perfectly balanced at the exact moderate point and the maximal $D_h$ always takes the largest possible value of 4.1869; above $T_c$, with increasing temperature, the entropy contribution becomes larger and larger than the enthalpy contribution, and the maximal $D_h$ at the general moderate point becomes smaller and smaller.

The probability density distributions $f(h)$ at an arbitrarily chosen concentration $L = 15$ nm and various temperatures are plotted in Figure 2a. According to Figure 1, the moderate-point temperature at this concentration is $T_m = 296.5$ K. At the low temperatures of 210 K and 250 K, due to the dominated enthalpy contribution, polyglutamine molecules aggregate tightly, leading to larger probabilities for larger $h$ values. In contrast, smaller $h$ values have larger probabilities at the high temperatures of 350 K and 390 K when the entropy effect dominates. In between, at the moderate point $T_m = 296.5$ K, the distribution of $f(h)$ is almost symmetric, resulting in the largest HOP fluctuation near this point.

For a fixed temperature, varying $L$ can also change the balance between entropy and enthalpy. The distributions at an arbitrarily selected $T = 360$ K and various concentrations are plotted in Figure 2b. According to Figure 1, the corresponding moderate-point box size $L_m$ = 12.435 nm, at which $f(h)$ is broader than the distributions at any other concentrations by satisfying Eq. (10), although it is still asymmetric.

The response function $\chi$ corresponding to $h$ has been calculated according to Eq. (15) and are shown in Figure 3. As shown in Figure 3a, $\chi = \beta D_h$ at $L$ = 15 nm also exhibits a peak, but the corresponding temperature is 292.7 K, slightly different from $T_m$ = 296.5 K due to the prefactor $\beta$. On the other hand, it is not surprising that at $T$ = 360 K, $\chi$ reaches its maximum at the moderate point $L_m$ = 12.435 nm (Figure 3b), since the prefactor $\beta$ is now a fixed value. Maximization of the response function around the moderate point manifests that, in the vicinity of the moderate point, a small change in the generalized field conjugate to the order parameter results in a large change of the order parameter. In other words, the system is very sensitive to the change of thermodynamic conditions. Two examples are the sensitivity of peptide self-assembly to temperature change[24, 25] and the sensitivity of ionic liquid properties to molecular structure.[26, 27]

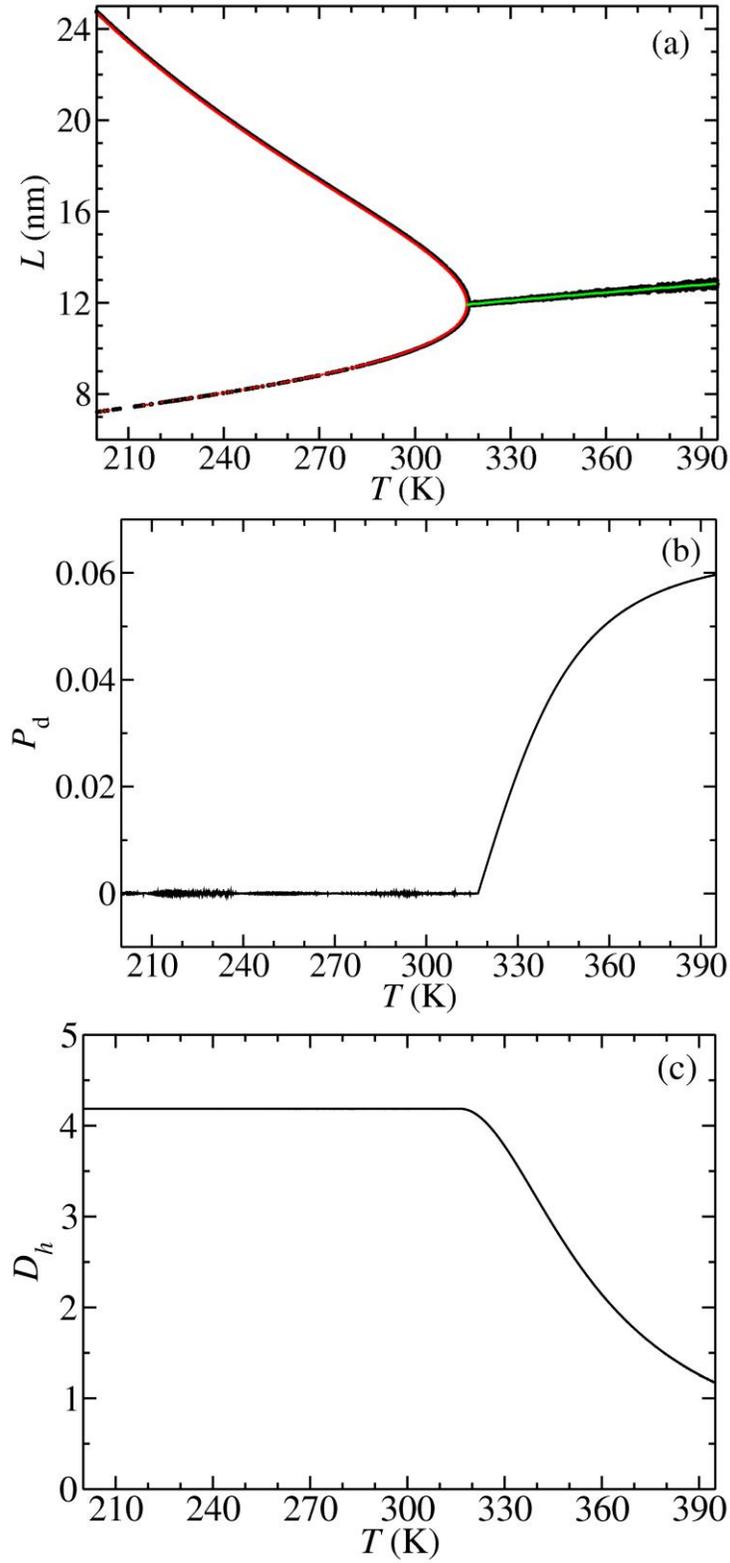

**Figure 1.** (a) Thermodynamic states with the largest fluctuation of the order parameter (red and green lines) and the moderate points determined by Eq. (10) (black lines). (b) Smallest probability difference defined in Eq. (10) as a function of temperature. (c) The largest order parameter fluctuation as a function of temperature.

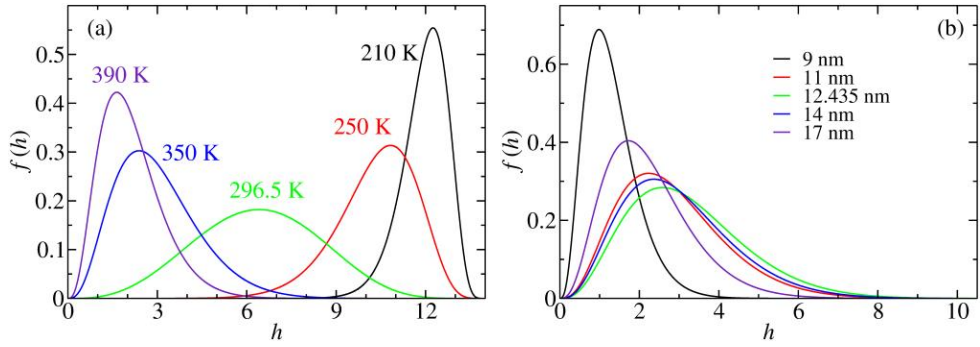

**Figure 2.** Probability density distributions at (a) $L = 15$ nm and various temperatures, and (b) $T = 360$ K and various concentrations.

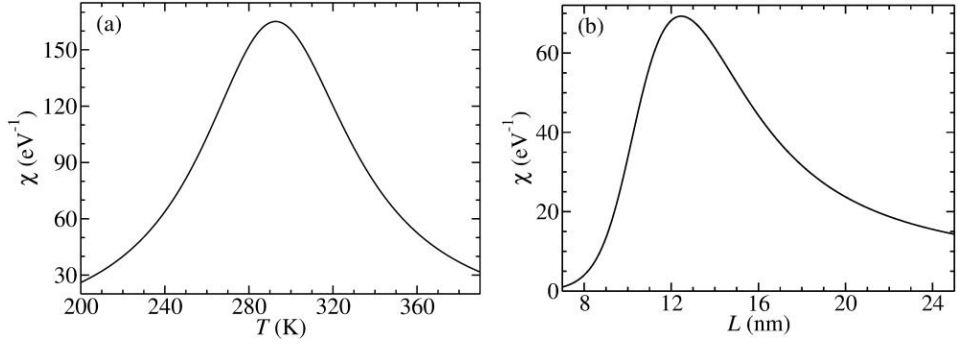

**Figure 3.** Response functions at (a) $L = 15$ nm as a function of temperature and (b) $T = 360$ K as a function of concentration.

## IV. EXAMPLES OF POSSIBLE APPLICATIONS

The moderation theory should be quite general and should be able to be applied to most (if not all) soft materials. A possible application of the two-substate model moderation theory could be the case of chemical bond breaking and reformation. If all microsates of a chemical bond can be grouped into only two substates independent of temperature: the enthalpy-dominated connected state and the entropy-dominated broken state, then at the moderate temperature given by Eq. (7), the chemical bond has equal probabilities of being connected or broken. Well above the moderate temperature, the bond is broken most of the time due to thermal fluctuation, and is connected most the time well below the moderate temperature. This simplified picture might be helpful for understanding the functions of hydrogen bonds in biological systems.

Another possible application is the determination of the best working temperature of biochannels. A biochannel works the best when the difference between the open state and the close state, modulated by a binary variable (e.g., binding and unbinding of an ion on a specific binding site) $B = B_0$ or $B_1$, is the largest. If we choose the width of the channel to be the order parameter $h$, the difference is

$$\langle \Delta h(\beta) \rangle = \langle h \rangle(\beta, B_1) - \langle h \rangle(\beta, B_0). \tag{24}$$

If the system energy $U$ linearly depends on $h$, the maximization requirement $\frac{\partial}{\partial \beta}\langle \Delta h(\beta) \rangle = 0$ leads to the condition at the moderate point

$$D_h(\beta_m, B_1) = D_h(\beta_m, B_0). \tag{25}$$

Alternatively, integrating the first part of Eq. (15) leads to

$$\Delta h(\beta) = \int_{B_0}^{B_1} \chi(\beta, B) dB. \tag{26}$$

Along with the second part of Eq. (15), we obtain

$$\beta_m = -\frac{\int_{B_0}^{B_1} D_h(\beta, B) dB}{\int_{B_0}^{B_1} \frac{\partial}{\partial \beta} D_h(\beta, B) dB}. \tag{27}$$

Either Eq. (25) or (27) can be used to determine the best working temperature of a biochannel.

**V. CONCLUSIONS AND DISCUSSION**

In summary, we developed the moderation theory for soft matter by defining the moderate point of a statistical system at which the entropy and enthalpy contributions among substates along an order parameter perfectly balance each other or have a minimal difference. The order parameter fluctuation maximizes around the moderate point. Soft materials work in the vicinity of the moderate point, which explains their sensitivity to thermodynamic condition changes via the relation between the order parameter fluctuation and the associated response function. The mesoscopic characteristic spatial and temporal scales of soft matter are also explained by the maximization of the finite spatial correlation length at the moderate point. The moderation theory was validated by the simplified statistical model for polyglutamine aggregation, and its possible applications to switching chemical bonds and biochannels were also briefly discussed. The moderation theory is anticipated to form the basis of the theoretical framework providing a quantitative definition for soft matter.

An important feature of the moderation theory is that the moderate point heavily depends on the order parameter because a specific choice of an order parameter reflects the perspective of the study for a given soft material. In addition, it is worth emphasizing that, the maximization of the response function is not always exactly at the moderate point, and there are no distinct boundaries for the entropy and enthalpy difference dividing the "soft" and "non-soft" regions of the system, reflecting the "softness" feature of soft matter.

More questions related to the moderation theory for soft matter are open for investigation. Figure 1 demonstrates that the polyglutamine aggregation model resembles a liquid-gas phase transition of a finite-size system, and the line above the critical temperature is very likely the so-called Widom line.[28-30] Therefore, the relations between the moderation theory and the phase transition of a finite-size system as well as the Widom line are worth investigating. In addition, it is interesting to see how the current terminologies of entropy-enthalpy compensation and entropy-enthalpy balance fit in the moderation theory. Finally, the moderation theory might help to explain why soft matter can self-assemble into ordered structures, a unique feature of soft matter very important in various scientific areas including physics, chemistry, biology, and materials science.

## ACKNOWLEDGMENTS

We thank Rui Shi and Sen Zhou for delightful discussions and critical readings of the manuscript. This work was supported by the National Basic Research Program of China (973 program, No. 2013CB932804) and the National Natural Science Foundation of China (Nos. 11274319 and 11421063). Allocations of computer time from the SCCAS and the HPC Cluster of ITP-CAS are gratefully acknowledged.## APPENDIX

### APPENDIX A: MAXIMIZATION OF ORDER PARAMETER FLUCTUATION AT MODERATE POINTS FOR THE TWO-SUBSTATE MODEL

For the two-substate model, since the free energies

$$F_i = -\frac{1}{\beta}\ln Z_i = -\frac{1}{\beta}\left(\ln P_i + \ln Z\right) \tag{A1}$$

with $P_i$ the probability and $Z_i$ the partial partition function for substate $i$, $i$=1 or 2, and $Z$ the total partition function, when assuming $P_1 \geq P_2$ (i.e., $F_1 \leq F_2$), the moderate condition

$$|F_1 - F_2| = F_d \tag{A2}$$

is equivalent to

$$P_1 / P_2 = c_P, \tag{A3}$$

where $c_P \equiv \exp(\beta F_d) \geq 1$ is the smallest possible ratio between $P_1$ and $P_2$. Along with the normalizing condition $P_1 + P_2 = 1$, we have

$$P_1 = \frac{c_P}{1+c_P} \text{ and } P_2 = \frac{1}{1+c_P}. \tag{A4}$$

The order parameter fluctuation

$$D_h \equiv \langle h^2 \rangle - \langle h \rangle^2 = \frac{c_P}{(1+c_P)^2}(h_1 - h_2)^2. \tag{A5}$$

When varying temperature $T$, one can always find a moderate temperature $T_m = \frac{U_2 - U_1}{k_B (\ln g_2 - \ln g_1)}$ with $g_i$ and $U_i$, $i$=1 or 2 the density of states and total energy, respectively, to have $c_P = 1$, which leads to the largest possible $D_h$, namely, $D_h$ maximizes at the exact moderate point. When varying the generalized field $B$, sometimes the smallest possible $c_P$ is larger than 1. In this case, let $D'_h$ be the order parameter fluctuation corresponding to an arbitrary $c'_P = c_P + x$ with $x > 0$, then

$$D'_h - D_h = \frac{x(1 - xc_P - c_P^2)}{(1+c_P+x)^2 (1+c_P)^2}(h_1 - h_2)^2 < 0. \tag{A6}$$

Therefore, $D_h$ also maximizes at the general moderate point when varying the generalized field $B$. Since the expression of $D_h$ does not change when switching $h_1$ and $h_2$, it is obvious that the same conclusion is also valid for the case $P_1 \leq P_2$.

**APPENDIX B: MODERATE CONDITIONS FOR THE CONTINUOUS CASE**

We map a continuous case to the two-substate model by dividing the continuous probability density function $f(h) \equiv g(h)\exp(-\beta U(h))/Z$ into two parts at the expected value of the order parameter $\langle h \rangle$, and regard these two parts as the entropy-dominated state 1 and enthalpy-dominated state 2 in the two-substate model, whose partial partition

functions

$$Z_1 \equiv Z\int_{-\infty}^{\langle h \rangle} f(h)\mathrm{d}h \text{ and } Z_2 \equiv Z\int_{\langle h \rangle}^{+\infty} f(h)\mathrm{d}h .$$ (B1)

According to the exact moderate condition for the two-substate model $F_1 = F_2$, the exact moderate condition for the continuous case is then

$$\int_{-\infty}^{\langle h \rangle} f(h)\mathrm{d}h = \int_{\langle h \rangle}^{+\infty} f(h)\mathrm{d}h .$$ (B2)

Since the definition of the median $h_\mathrm{m}$ is

$$\int_{-\infty}^{h_\mathrm{m}} f(h)\mathrm{d}h = \int_{h_\mathrm{m}}^{+\infty} f(h)\mathrm{d}h = 0.5 ,$$ (B3)

the above condition is equivalent to

$$\langle h \rangle = h_\mathrm{m} .$$ (B4)

The general moderate condition for the continuous case can also be derived from the condition for the two-substate model given by Eq. (A2) as follows. For convenience, first we still assume $F_1 \leq F_2$, then for the continuous case, Eq. (A2) leads to

$$\frac{Z_1}{Z_2} = \exp(-\beta F_\mathrm{d}) \equiv c_\mathrm{P} \geq 1 .$$ (B5)

Utilizing the definition of the median Eq. (B3) and define

$$R \equiv \int_{h_\mathrm{m}}^{\langle h \rangle} f(h)\mathrm{d}h ,$$ (B6)

Eq. (B5) can be written as

$$\frac{0.5 + R}{0.5 - R} = c_\mathrm{P} .$$ (B7)

Therefore, we have the general moderate condition for the continuous case

$$R \equiv \int_{h_\mathrm{m}}^{\langle h \rangle} f(h)\mathrm{d}h = P_\mathrm{d} ,$$ (B8)

where $P_\mathrm{d} \equiv 0.5 - \frac{1}{1+c_\mathrm{P}} \geq 0$ has the smallest possible value. When $c_\mathrm{P} \equiv \frac{Z_1}{Z_2} = 1$, the above condition reduces to the exact moderate condition Eq. (B4). To allow Eq. (B8) also holds for the case $F_1 \geq F_2$, it should be generalized to

$$|R| \equiv \left| \int_{h_m}^{\langle h \rangle} f(h) \mathrm{d}h \right| = P_d \tag{B9}$$

with $P_d \geq 0$ the smallest possible value, which is the general moderate condition for the continuous case.

## APPENDIX C: CONDITION FOR THE MAXIMIZATION OF ORDER PARAMETER FLUCTUATION FOR THE CONTINUOUS CASE

The order parameter fluctuation is generally a function of temperature and generalized field $D_h(\beta, B)$. When fixing the temperature, the generalized field value $B_m$ maximizing $D_h(\beta, B)$ must satisfy

$$\left( \frac{\partial D_h}{\partial B} \right)_\beta = 0. \tag{C1}$$

In the linear response regime, the system Hamiltonian

$$U(h) = Bh - U_0(h), \tag{C2}$$

so the system partition function

$$Z = \int_{-\infty}^{+\infty} g(h) \exp(\beta B h - \beta U_0) \mathrm{d}h. \tag{C3}$$

Put $D_h \equiv \langle h^2 \rangle - \langle h \rangle^2$ into Eq. (C1) and utilizing Eq. (C3), we obtain the condition for $D_h$ to maximize with respect to $B$

$$\left( \frac{\partial D_h}{\partial B} \right)_\beta = \beta \left( \langle h^3 \rangle - 3 \langle h^2 \rangle \langle h \rangle + 2 \langle h \rangle^3 \right) = \beta \left\langle (h - \langle h \rangle)^3 \right\rangle = 0. \tag{C4}$$

If the system energy linearly depends on the order parameter $U = ah$ with $a$ a constant, the same mathematical procedure also leads to

$$\left( \frac{\partial D_h}{\partial \beta} \right)_B = a \left\langle (h - \langle h \rangle)^3 \right\rangle = 0. \tag{C5}$$

Therefore, the requirement for $D_h$ to maximize is equivalent to the condition

$$\left\langle (h - \langle h \rangle)^3 \right\rangle = 0. \tag{C6}$$

**APPENDIX D: MAXIMIZATION OF ORDER PARAMETER FLUCTUATION AT MODERATE POINTS FOR THE CONTINUOUS CASE**

Let $P_1 \equiv R + 0.5 = \int_{-\infty}^{\langle h \rangle} f(h) dh$. When $P_1 > 0.5$, the distribution $f(h)$ has a positive skewness, so $\langle (h - \langle h \rangle)^3 \rangle > 0$; at the same time, a larger $P_1$, i.e., a more positively skewed distribution, has a more left-shifted mean value $\langle h \rangle$, so $\left( \frac{\partial \langle h \rangle}{\partial P_1} \right)_\beta < 0$. Therefore, when the temperature is fixed, the order parameter fluctuation $D_h$ changes with $P_1$ as

$$\left( \frac{\partial D_h}{\partial P_1} \right)_\beta = \left( \frac{\partial D_h}{\partial B} \right)_\beta \left( \frac{\partial B}{\partial \langle h \rangle} \right)_\beta \left( \frac{\partial \langle h \rangle}{\partial P_1} \right)_\beta = \frac{\langle (h - \langle h \rangle)^3 \rangle}{D_h} \left( \frac{\partial \langle h \rangle}{\partial P_1} \right)_\beta < 0. \tag{D1}$$

That is, when $P_1 > 0.5$ ($R > 0$), $D_h$ monotonically decreases with increasing $P_1$.

When $P_1 < 0.5$, $f(h)$ has a negative skewness, so $\langle (h - \langle h \rangle)^3 \rangle < 0$; at the same time, a larger $P_1$, i.e., a less negatively skewed distribution, has a more right-shifted mean value $\langle h \rangle$, so we still have $\left( \frac{\partial \langle h \rangle}{\partial P_1} \right)_\beta < 0$, and

$$\left( \frac{\partial D_h}{\partial P_1} \right)_\beta = \frac{\langle (h - \langle h \rangle)^3 \rangle}{D_h} \left( \frac{\partial \langle h \rangle}{\partial P_1} \right)_\beta > 0. \tag{D2}$$

That is, when $P_1 < 0.5$ ($R < 0$), $D_h$ monotonically increases with increasing $P_1$.

When $P_1 = 0.5$ ($R = 0$), $\langle h \rangle = h_m$, $f(h)$ has a zero skewness, the system is at the exact moderate point and $D_h$ reaches its maximal value. For the general case satisfying Eq. (B9), since $D_h$ monotonically decreases with $|R|$, at the general moderate point, the system still has an order parameter fluctuation $D_h$ as large as possible.

The above is also true when fixing $B$ and varying $\beta$ under the assumption that the system energy $U(h)$ linearly depends on $h$. Note that the above arguments are based on the fact that $f(h)$ has the mathematical form of a Boltzmann distribution and not necessarily true for an arbitrary distribution.

**APPENDIX E: MAXIMIZATION OF RESPONSE FUNCTION**

In the linear response regime,

$$f(h;\beta,B) = g(h)\exp(\beta Bh - \beta U_0)/Z(\beta,B), \tag{E1}$$

where $Z(\beta,B) = \int_{-\infty}^{+\infty} f(h;\beta,B)dh$. Since the system free energy $F = -\frac{1}{\beta}\ln Z$, its derivative with respect to $B$ is

$$\left(\frac{\partial F}{\partial B}\right)_\beta = -\frac{1}{\beta Z}\left(\frac{\partial Z}{\partial B}\right)_\beta = -\langle h \rangle. \tag{E2}$$

The corresponding response function

$$\chi(\beta,B) \equiv \left(\frac{\partial \langle h \rangle}{\partial B}\right)_\beta = \beta\left(\langle h^2 \rangle - \langle h \rangle^2\right) = \beta D_h(\beta,B). \tag{E3}$$

Therefore, when $\beta$ is fixed and $B$ varies, the response function $\chi$ maximizes exactly at the point when $D_h$ maximizes. However, when $B$ is fixed and $\beta$ varies, $\chi$ does not maximize at the point $D_h$ maximizes. Below we will show that the difference between the points for $\chi$ and $D_h$ maximize when varying $\beta$ is small if $D_h$ changes slowly around its maximization point.

Suppose $\chi(\beta)$ maximizes at the point $\beta_0$, we have

$$\left(\frac{\partial \chi}{\partial \beta}\right)_B(\beta_0) = D_h(\beta_0) + \beta_0\left(\frac{\partial D_h}{\partial \beta}\right)_B(\beta_0) = 0$$
$$\Rightarrow \left(\frac{\partial D_h}{\partial \beta}\right)_B(\beta_0) = -\frac{D_h(\beta_0)}{\beta_0} \tag{E4}$$

On the other hand, suppose $D(\beta)$ maximizes at another point $\beta_1: \left(\frac{\partial D_h}{\partial \beta}\right)_B(\beta_1) = 0$, and $\beta_1$ is not far from $\beta_0$. By doing the Taylor expansion for $D_h(\beta)$ near $\beta_0$, we have

$$D_h(\beta_1) = D_h(\beta_0) + \left(\frac{\partial D_h}{\partial \beta}\right)_B(\beta_0)(\beta_1 - \beta_0) + \Delta\left((\beta_1 - \beta_0)^2\right)$$
$$\Rightarrow D_h(\beta_1) - D_h(\beta_0) \approx \left(\frac{\partial D_h}{\partial \beta}\right)_B(\beta_0)(\beta_1 - \beta_0) = -\frac{D_h(\beta_0)}{\beta_0}(\beta_1 - \beta_0) \tag{E5}$$

Therefore, the relative deviation of the point for maximization is

$$\frac{\beta_1 - \beta_0}{\beta_0} \approx \frac{D_h(\beta_0) - D_h(\beta_1)}{D_h(\beta_0)}. \tag{E6}$$

The above equation indicates that, the deviation of the temperature at which $\chi$ maximizes away from the temperature at which $D_h$ maximizes is small if $D_h$ varies slowly around its maximization temperature.

**APPENDIX F: MAXIMIZATION OF SPATIAL CORRELATION LENGTH**

In the linear response regime, the response function can be connected to the spatial correlation by

$$\frac{1}{\beta}\chi(r) \sim \langle \delta h(0) \delta h(r) \rangle, \tag{F1}$$

where $r$ is the relative distance between two locations, $\chi(r)$ is the partial response function with $\chi = \int_0^{+\infty} \chi(r) dr$, and $\delta h(r) \equiv h(r) - \langle h \rangle$ is the local fluctuation of the order parameter $h$. If we denote the length of the characteristic spatial scale as $\xi$, then we approximately have $\langle \delta h(0) \delta h(r) \rangle \sim \exp(-r/\xi)$, and so

$$\int_0^{+\infty} \frac{1}{\beta} \chi(r) dr \sim \int_0^{+\infty} \exp(-r/\xi) dr \Rightarrow D_h = \frac{1}{\beta}\chi \sim \xi, \tag{F2}$$

which means that the characteristic spatial correlation length maximizes at the moderate point when $D_h$ maximizes.